\documentclass[sigconf, screen]{acmart}

\usepackage{array}
\usepackage{graphicx} 
\usepackage[skip=1pt]{caption}
\usepackage{subcaption} 
\usepackage[inline]{enumitem}

\usepackage{bbm}
\usepackage{nameref,hyperref}
\PassOptionsToPackage{table}{xcolor}

\usepackage[normalem]{ulem}

\usepackage{csquotes}

\usepackage{listings}

\usepackage{acronym}
\usepackage{tabularx}
\usepackage{longtable}
\usepackage{tcolorbox}
\usepackage{xcolor}
\AtBeginDocument{%
  \providecommand\BibTeX{{%
    \normalfont B\kern-0.5em{\scshape i\kern-0.25em b}\kern-0.8em\TeX}}}

\acrodef{GNN}{graph neural network}
\acrodef{LBSN}{location-based social network}
\acrodef{LLM}{large language model}
\acrodef{POI}{point-of-interest}
\acrodef{RNN}{recurrent neural network}
\acrodef{PEFT}{parameter-efficient fine-tuning}
\acrodef{NF}{NormalFloat}
\acrodef{FP}{floating point}
\acrodef{BF}{BrainFloating}
\acrodef{LoRA}{low-rank adaptation}
\acrodef{RFT}{reinforcement fine-tuning}
\acrodef{SFT}{supervised fine-tuning}
\acrodef{RL}{reinforcement learning}
\acrodef{MRR}{mean reciprocal rank}
\acrodef{RR}{reciprocal rank}
\acrodef{SID}{semantic ID}
\acrodef{SOM}{self organizing map}
\acrodef{RSOM}{residual self organizing map}
\acrodef{NICC}{normalized intra-class compactness}
\acrodef{NICS}{normalized inter-class separation}

\setcopyright{rightsretained}
\acmISBN{}
\acmDOI{}

\author{Peibo Li}
\orcid{0009-0001-0201-3564} 
\affiliation{%
  \institution{University of New South Wales}
  \city{Sydney}
  \country{Australia}
}
\email{peibo.li@student.unsw.edu.au}

\author{Shuang Ao}
\orcid{0009-0008-3321-8416}
\affiliation{%
  \institution{University of New South Wales}
  \city{Sydney}
  \country{Australia}
}
\email{shuang.ao@unsw.edu.au}

\author{Hao Xue}
\orcid{0000-0003-1700-9215} 
\affiliation{%
  \institution{The Hong Kong University of Science and Technology (Guangzhou)}
  \city{Guangzhou}
  \country{China}
}
\email{haoxue@hkust-gz.edu.cn}

\author{Yang Song}
\orcid{0000-0003-1283-1672}
\affiliation{%
  \institution{University of New South Wales}
  \city{Sydney}
  \country{Australia}
}
\email{yang.song1@unsw.edu.au}
\author{Maarten de Rijke}
\orcid{0000-0002-1086-0202} 
\affiliation{%
  \institution{University of Amsterdam}
  \city{}
  \country{The Netherlands}
}
\email{m.derijke@uva.nl}

\author{Johan Barthélemy}
\orcid{0000-0002-7800-5309}
\affiliation{%
  \institution{Nvidia}
  \country{USA}
}
\email{jbarthelemy@nvidia.com}

\author{Tomasz Bednarz}
\orcid{0000-0001-9240-0922}
\affiliation{%
  \institution{Nvidia}
  \country{USA}
}
\email{tbednarz@nvidia.com}

\author{Flora D. Salim}
\orcid{0000-0002-1237-1664}
\affiliation{%
  \institution{University of New South Wales}
  \city{Sydney}
  \country{Australia}
}
\email{flora.salim@unsw.edu.au}

\renewcommand{\shortauthors}{Peibo Li et al.}



\title[Refine-POI: Reinforcement Fine-Tuned Large Language Models for Next Point-of-Interest Recommendation]{\texorpdfstring{Refine-POI: Reinforcement Fine-Tuned Large Language Models\\ for Next Point-of-Interest Recommendation}{Refine-POI: Reinforcement Fine-Tuned Large Language Models for Next Point-of-Interest Recommendation}}

\begin{document}

\begin{abstract}
Advancing \acp{LLM} for the next \ac{POI} recommendation task faces two fundamental challenges. (i)~While existing methods produce \aclp{SID} that incorporate semantic information, their topology-blind indexing fails to preserve semantic continuity, meaning that proximity in ID values mirrors the coherence of the underlying semantics. And (ii)~\ac{SFT}-based methods restrict model outputs to top-1 predictions. These approaches suffer from ``answer fixation'' and, owing to the scarcity of supervision, produce neither ranked top-$k$ lists nor reasoning that explains a recommendation. We propose Refine-POI, a framework that addresses these challenges through topology-aware ID generation and \acl{RFT}. First, we introduce a hierarchical \acl{SOM} quantization strategy to generate \aclp{SID}, ensuring that coordinate proximity in the codebook reflects semantic similarity in the latent space. Second, we employ a policy gradient framework to optimize the generation of top-$k$ recommendation lists, liberating the model from strict label matching. Extensive experiments on three real-world datasets demonstrate that Refine-POI significantly outperforms state-of-the-art baselines. Furthermore, we identify a reasoning-degradation failure mode in which naive \ac{RFT} drives the model's reasoning from grounded to generic, trace its root cause to entropy collapse during training, and show that a lightweight entropy regularizer restores grounded reasoning at no accuracy cost. The source code is available at \url{https://github.com/neolifer/RefinePOI/}
\end{abstract}

\maketitle

\acresetall

\section{Introduction}
\begin{figure*}[t]
    \centering
    \begin{subfigure}[b]{0.55\textwidth}
        \centering
        \begin{tabular}{ccc}
        \includegraphics[height=3.25cm]{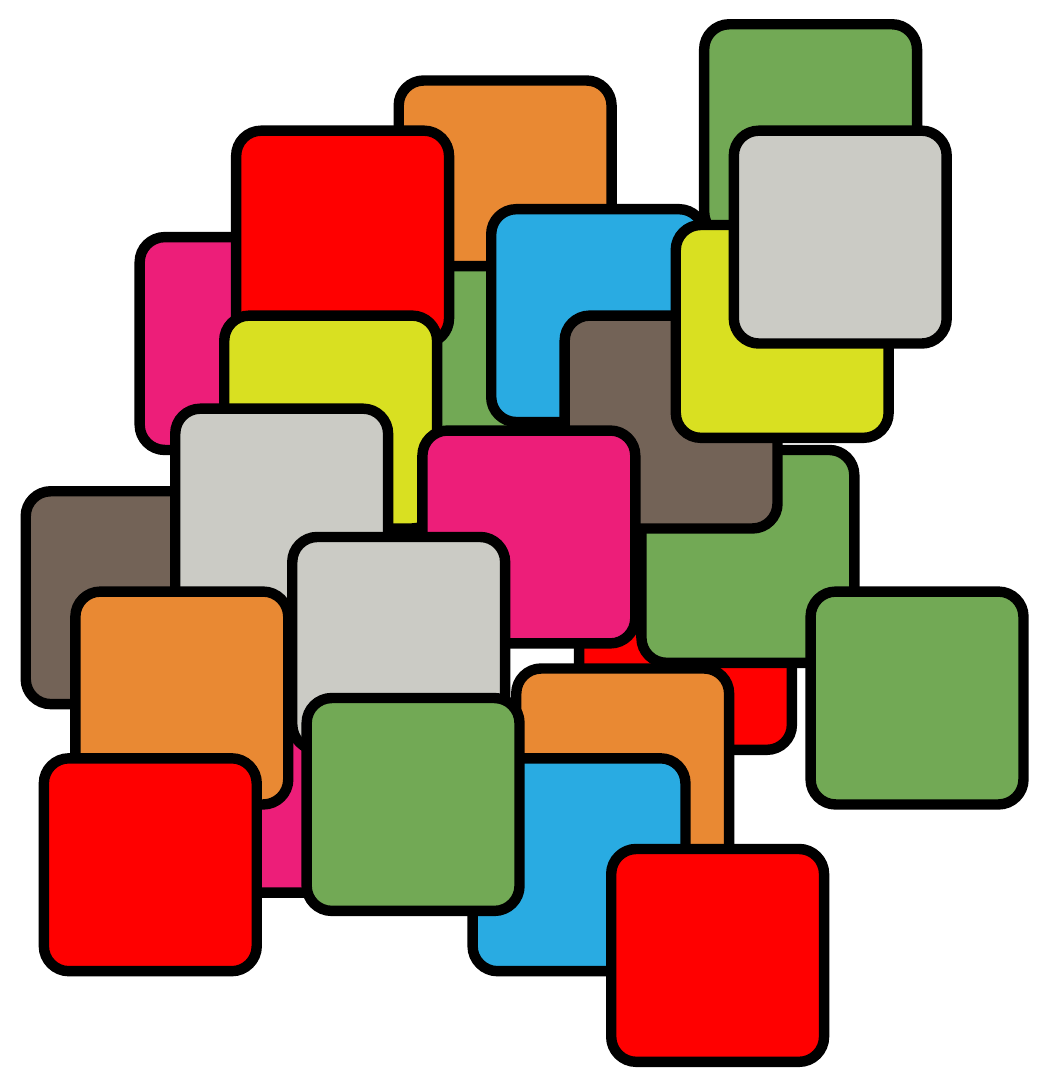}
        & \raisebox{1.5cm}{vs.} &
        \includegraphics[height=3.25cm]{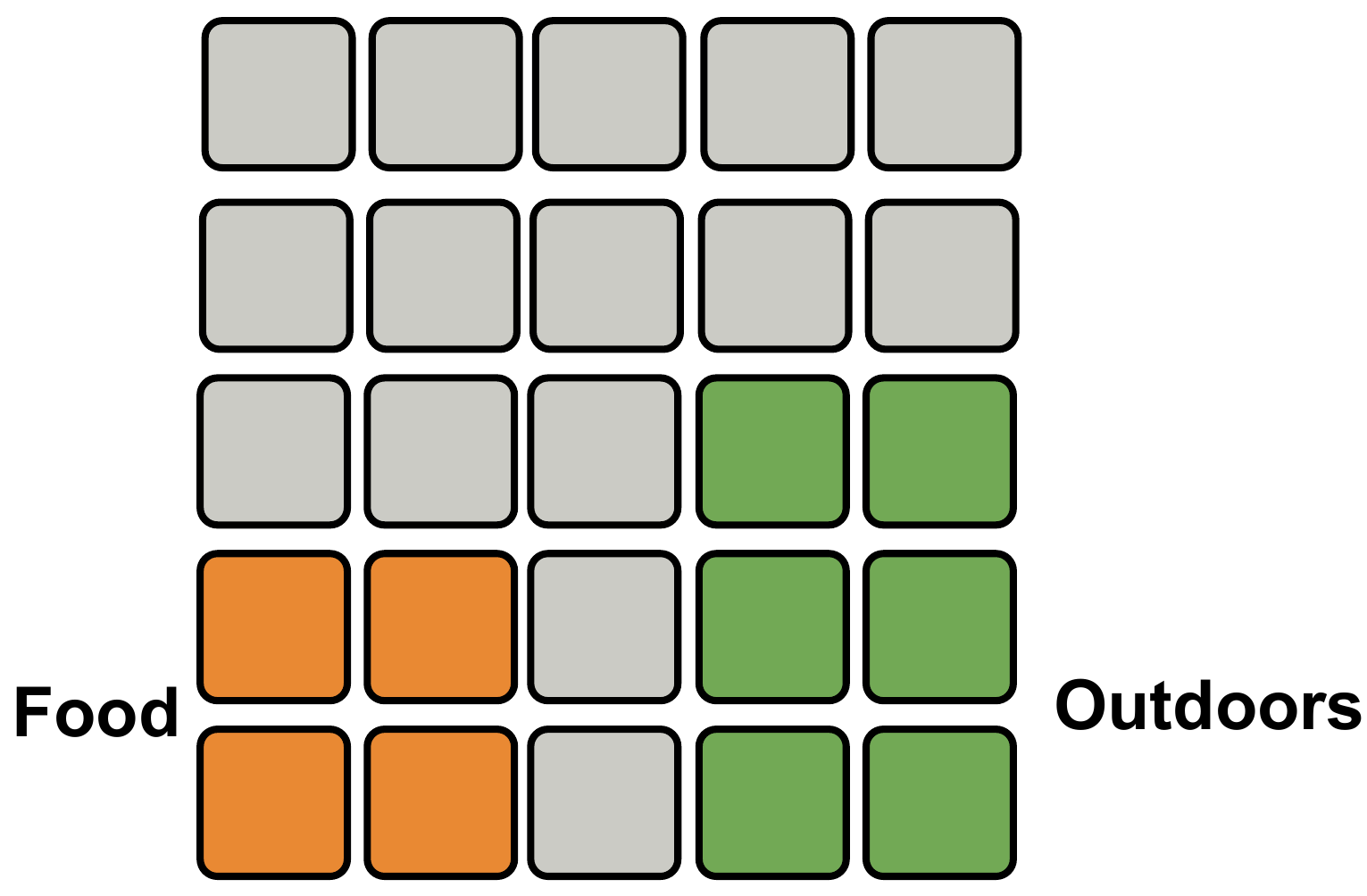}
        \end{tabular}
        \caption{(Left): Topology-blind indexed \acp{SID}. (Right): Topology-aware SIDs.}
        \label{fig:sub1}        
    \end{subfigure}
    \hfill 
    \begin{subfigure}[b]{0.4\textwidth}
        \centering
        \begin{tabular}{c}
        \includegraphics[width=5cm]{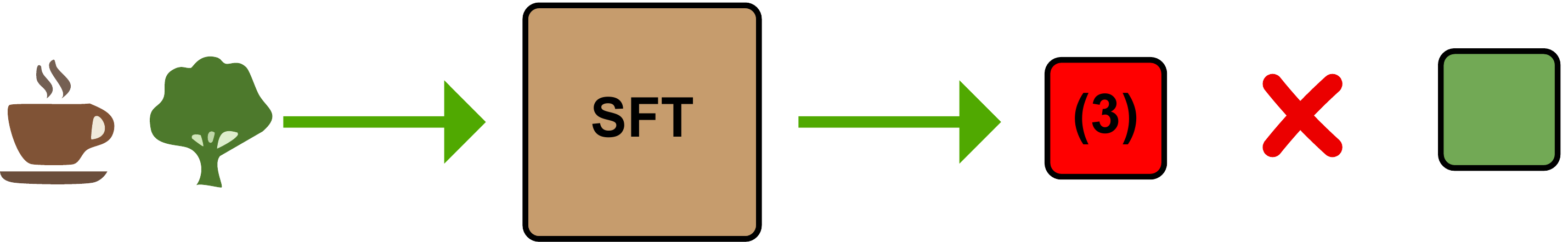}
        \\
        vs.
        \\
        \includegraphics[width=5cm]{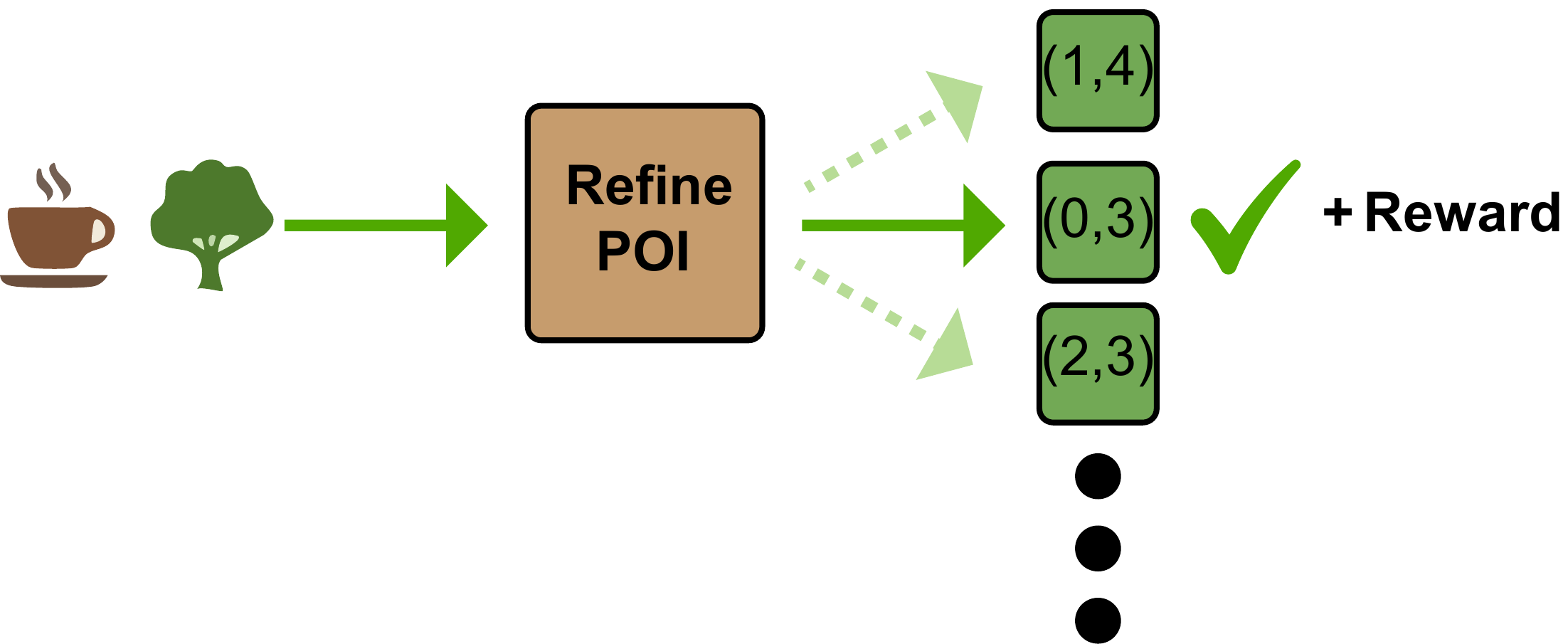}
        \end{tabular}
        \caption{(Top): Supervised fine-tuning. (Bottom): Reinforcement fine-tuning. \phantom{One}}
        \label{fig:sub2}
    \end{subfigure}
    
    
    \caption{The key differences between topology-blind \acp{SID} with \ac{SFT} and topology-aware \acp{SID} with Refine-POI. The \acp{SID} in (a, left) refer to existing methods that do not consider semantic continuity between close indices. \ac{SFT} (b, top) can only perform exact matching, so the model is trained to produce a single item. The \acp{SID} in (a, right) are constructed as coordinates in a map, where close coordinates are also similar in semantics. Refine-POI trains the model with \ac{RFT} (b, bottom); the model is encouraged to produce top-$k$ recommendation lists, and the correct item is rewarded based on its position.}
    \label{fig:first_figure}
\end{figure*}

Next \ac{POI} recommendation uses users' past check-in trajectories to predict their future mobility. It has recently shifted from a traditional sequence of graph models to \ac{LLM}-based approaches. 
Despite their progress, current \ac{LLM}-based approaches are constrained by two fundamental challenges at the representation level and the training level.

First, the \textit{representation challenge}. Traditional non-\ac{LLM} methods represent \acp{POI} as atomic IDs that contain no semantic information. Recent \ac{LLM}-based methods produce semantic representations of IDs called \emph{\aclp{SID}} (\acp{SID}) \cite{rajput2023recommender, mei2025semantic, 10.1145/3711896.3736981}, which are generated by mapping POI content information into codebook vectors. While these methods effectively introduce semantic information to IDs, they overlook the geometric properties of the representation. The codebook is essentially an unordered set of vectors. The mapping from raw data to discrete indices is based solely on Euclidean distance minimization and random initialization, leading to topology-blind indexing. Consequently, code vectors with adjacent indices may represent semantically disparate regions in the latent space. This topology-blind indexing hinders the model's ability to extract meaningful patterns from ID sequences.

Second, the challenge of \textit{task misalignment} arising from supervision scarcity. To adapt \ac{LLM}s for recommendation, existing methods typically fine-tune models using a QA format, where the training objective is to exactly match the provided \textit{single} ground-truth \ac{POI}. A critical gap exists between this limited supervision and the sophisticated capabilities required for an intelligent recommender. 

\vspace*{1.5mm}\noindent%
\textbf{Our research questions.}
Motivated by the challenges listed above, we pursue two research questions: 
\begin{enumerate}[label=(\arabic*),leftmargin=*]
\item \textbf{Can we generate \acp{SID} that preserve semantic continuity?} \item \textbf{Can we fine-tune an LLM to produce a top-$k$ recommendation list and reasoning process \emph{without} requiring extra ground-truth labels?} 
\end{enumerate}

\vspace*{1.5mm}\noindent%
\textbf{Our approach.}
To answer our research questions, we propose Refine-POI, a \ac{RFT} framework for next POI recommendation. See Figure~\ref{fig:first_figure}.
The motivation for the design of Refine-POI comes from the following observations and advancements.
Real-world \ac{POI} datasets inherently lack explicit supervision for \textit{reasoning paths} (e.g., the rationale behind a visit) or optimal \textit{ranked lists} (e.g., potential alternative candidates). Consequently, \acf{SFT} forces the model to treat the task as a deterministic retrieval of a single ID, leading to ``answer fixation.'' Without guidance from rich labels, the model fails to learn how to generate diverse top-$k$ lists or articulate the decision-making process, reducing the \ac{LLM} to a black-box predictor restricted by the scarcity of the dataset.
\ac{RFT} directly addresses these issues. Instead of constraining the model to mimic a single label, \ac{RFT} enables learning from reward signals that evaluate the quality of the list. This enables optimization not only for whether the correct POI is included, but also for its position in the ranking and the diversity of recommendations. In other words, where SFT can only teach a model to ``guess the next item,'' \ac{RFT} can teach it to ``construct a high-quality top-$k$ list''. 
Recent advances, such as R1-like LLMs~\cite{guo2025deepseek}, demonstrate that \ac{RFT} can enhance reasoning and problem-solving ability. More broadly, reinforcement learning provides a way to refine LLM outputs beyond the strict constraints of ground-truth data, offering a promising direction for improving recommendation quality while preserving the LLM’s pre-trained capabilities.

Refine-POI generates \acfp{SID} using a hierarchical \ac{SOM} \cite{kohonen2002self}, which encodes \acp{SID} as coordinates. Code vectors with close coordinates are updated together during training. Therefore, neighboring code vectors can represent similar semantics, thereby preserving the semantic continuity. Refine-POI also builds on the \ac{SFT} pipeline by transforming check-in data into textual prompts for LLMs, but crucially replaces label imitation with policy gradient optimization. Instead of forcing the model to reproduce a single ground-truth POI, we evaluate its outputs through rule-based rewards. This shift is key: \ac{RFT} enables learning from list-level signals, where success is measured not only by whether the ground-truth POI appears, but also by its rank position and the diversity of the recommendation list. In contrast to conventional binary accuracy rewards, our recommendation-driven rewards provide richer feedback that directly aligns with top-$k$ objectives. As a result, Refine-POI can generate complete ranked lists rather than isolated top-1 predictions. Furthermore, we find that naive \ac{RFT} degrades the model's reasoning, leading it to recite generic templates rather than engage with the user's data; we trace this to entropy collapse during training and show that a lightweight entropy regularizer restores grounded reasoning. Together, these components allow Refine-POI to achieve strong top-$k$ accuracy while keeping its reasoning grounded in the user's history, addressing the fundamental challenges of existing approaches.


\vspace*{1.5mm}\noindent%
\textbf{Contributions.} Our main contributions are as follows:
\begin{enumerate}[leftmargin=*,nosep]
    \item We present Refine-POI, the first \ac{RFT}-based LLM framework for next \ac{POI} recommendation. Unlike \ac{SFT}, which is constrained to top-1 prediction, Refine-POI uses list-level reward signals, making it inherently capable of optimizing full top-$k$ recommendation lists without requiring extra ground-truth labels.
    \item We propose topology-aware \acp{SID}, which both introduces semantic information to POI IDs and preserves the semantic continuity.
    \item We design novel recommendation-driven rewards that move beyond binary correctness. Our formulation considers whether the ground-truth POI is included, its position in the ranking, and the diversity of the list, directly aligning optimization with real-world recommendation objectives.
    \item We conduct extensive experiments on three real-world \aclp{LBSN} datasets and show that Refine-POI achieves state-of-the-art performance.
    \item We identify entropy collapse as the root cause of reasoning degradation under \ac{RFT} in the next \ac{POI} recommendation setting, and show that entropy regularization restores grounded reasoning.
\end{enumerate}

\section{Related Work}
\subsection{LLM-based next POI recommenders}

\subsubsection{Prompt-based}
LLM-Mob \cite{wang2023would} is a prompt-based framework that guides LLMs to predict users' next location and provide explanations. The authors give LLMs the trajectory of a user that is provided using the format (start time, day of week, duration, place ID) and ask the model to predict the next place ID of the target stay. They also provide instructions that guide the model to `think.'
LLM-Move \cite{10605522} focuses on the geographical influence and sequential transitions of user trajectories. The task is asking the model to choose the next POI in a candidate set, which consists of the ground truth and 100 randomly selected candidates. Pre-trained LLMs are fed with long-term and recent check-ins and prompted to consider four aspects: long-term preference, current preference, distance, and the categories of the user-preferred sequential transitions. HMP-LLM \cite{zhong2024hmp} uses trend, seasonal, and residual time series data, which is derived using decomposition techniques. The authors guide LLMs to consider historical patterns and seasonal cycles. AgentMove \cite{feng2024agentmove} adopts an approach with better system design. It first encodes long-term spatial-temporal memory as user profiles and statistical patterns; short-term memory is also reflected in the prompt as statistical patterns. The model uses real addresses from a search engine to align LLMs with real-world knowledge. Finally, it has a location transition graph to capture shared mobility patterns. Prompt-based LLM recommenders have limited knowledge of data, as they can only access the data that appears in the prompt. Therefore, it is hard for them to recommend unseen \Acp{POI}. 

Our method is an \Ac{RFT}-based model, allowing the model to memorize \Acp{POI} throughout the training.

\subsubsection{Fine-tuning-based}
LLM4POI \cite{li2024large} transforms the next \ac{POI} recommendation task into question-answering. It applies \ac{SFT} with LLMs, focusing on using the rich contextual information in \ac{POI} data.
NextLocLLM \cite{liu2024nextlocllm} encodes locations based on continuous spatial coordinates and incorporates natural language descriptions of POI categories. To obtain top-\(k\) recommendations, the authors apply a KD-tree to retrieve the top-\(k\) locations closest to the predicted coordinates, which are the output from the LLM.
Genup \cite{wongso2024genup} generates user profiles and personas as a system prompt for \ac{SFT}, reducing reliance on extensive historical data and improving computational efficiency.  
GNPR-SID \cite{10.1145/3711896.3736981} generates \aclp{SID}, introducing semantic information to \ac{POI} IDs.
ComaPOI \cite{zhong2025comapoi} proposes a multi-agent framework that employs a match-and-rerank paradigm.
The biggest limitation of \ac{SFT}-based LLM recommenders is that the ground-truth they rely on does not contain a top-\(k\) recommendation list. As a result, they do not produce native top-\(k\) recommendations. 

Our \ac{RFT} framework is not constrained by the ground-truth by using reward signals; it can output native top-\(k\) recommendations.

\subsection{Reinforcement learning optimized LLMs in recommender systems}
There are various methods that use LLMs as reward models for reinforcement learning. Here, we only show the literature that uses LLMs as a policy.
LLMCRS \cite{feng2023large} uses reinforcement learning from performance feedback on conversational recommender systems (RLPF); it uses recommendation performance and response generation performance to guide LLM learning, resulting in improved overall performance for conversational recommenders. The authors use HIT and BLEU from LLMs' generated output as rewards to guide the training, and they use the basic policy gradient.
LLM2ER \cite{yang2024fine} uses an approach similar to RLHF to train LLMs to specifically explain recommendations. Instead of iteratively using human-generated data to train a reward model, the authors use a static dataset that consists of high-quality explanations.

Our \ac{RFT} uses recommendation-driven rewards, which do not rely on human feedback.

\section{Problem Definition}
The next POI recommendation problem can be formalized as follows. 
Consider a dataset \( \mathcal{D} \) of user check-in records. 
Each record is represented by a tuple \(q =  (u, p, c, t, g) \), where:
\begin{itemize}[leftmargin=*]
    \item \( u \) denotes a user from the set \( U = \{u_1, u_2, \ldots, u_N\} \), where \( N \) is the total number of users;
    \item \( p \) is a point of interest (POI) from the set $ P = \{p_1$, $p_2$, \ldots, $p_M\}$, where \( M \) is the number of distinct POIs;
    \item \( c \) specifies the category of the POI;
    \item \( t \) represents the timestamp of the check-in; and
    \item \( g \) signifies the geometric coordinate of the POI.
\end{itemize}

\noindent%
Given a time interval \( \Delta t \), trajectories for a user \( u \) are formed by splitting the check-in records based on this interval. 
Each trajectory \( T_i^u \) up to timestamp \( t \) for user \( u \) is given by:
\[ 
T_i^u(t) = \{(p_1, c_1, t_1, g_1), \ldots, (p_k, c_k, t_k, g_k)\}, 
\]
where \( t_1 < t_2 < \cdots < t_k = t \) and \( t_{k} - t_1 \leq \Delta t \).

Given a set of historical trajectories \( \mathcal{T}_u = \{T_1^u, T_2^u, \ldots, T_L^u\} \) for user \( u \), where \( L \) represents the number of trajectories for \( u \), the objective is to predict the POI \( p_{k+1} \)  for a new trajectory \( {T'}_i^u(t) \), where user \( u \) will check in at the immediate subsequent timestamp \( t_{k+1} \).

\begin{figure*}[h]
  \centering
  \includegraphics[width=\linewidth]{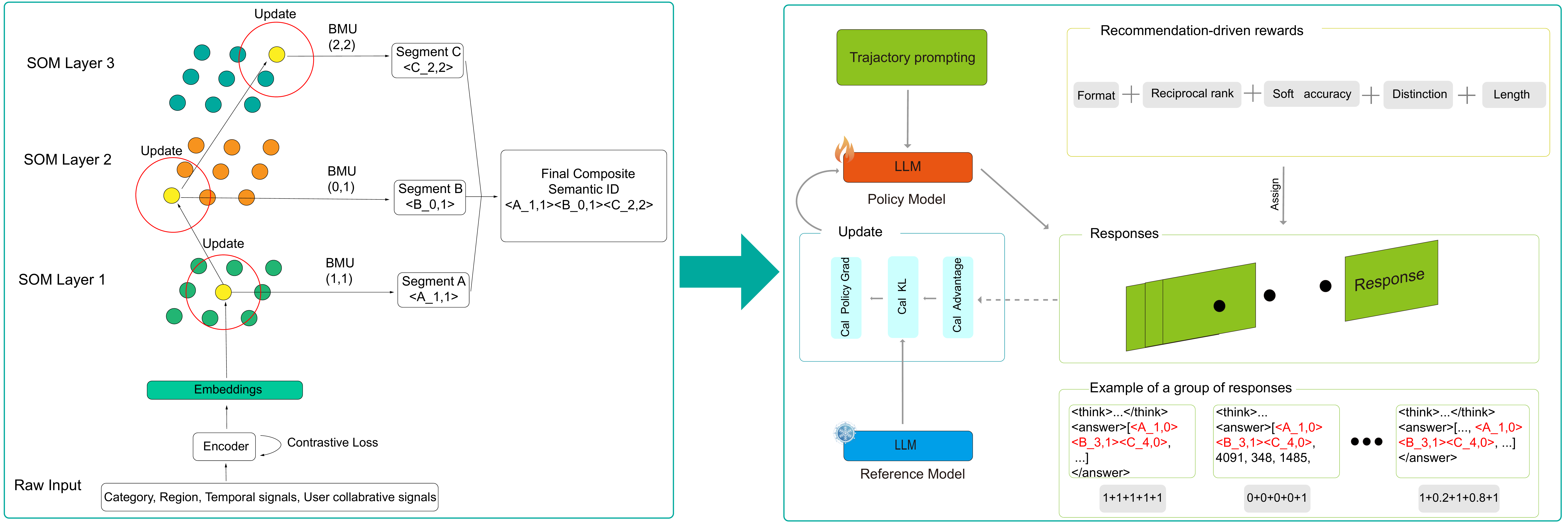}
  \caption{Overview of the Refine-POI framework, which consists of two module. (i) The left side is the topology-aware \ac{SID} generation with a hierarchical \ac{SOM}. And (ii) the right side is the \ac{RFT}. For training, we begin with trajectory prompting, in which we transform check-in records into prompts and enrich them with additional POI address information. Then we adopt GRPO \cite{shao2024deepseekmath} as the \ac{RFT} algorithm. In the bottom-right example, the first response has the correct format, the correct item appears in the first position, and all items are distinct. Thus, if we assume the output length is higher than the target length, the reward would be \(1\) (format reward) + \(\frac{1}{1}\) ( reciprocal rank reward) + \(1\) (soft accuracy reward) + \(10\times0.1\) (distinction reward) + \(1\) (length reward). The second response has the wrong format, despite containing the correct item. Therefore, the reward for the second response is 0. The last response has the correct format. The correct item appears in the 5-th position, and there are only eight distinct items. So the reciprocal rank reward and the distinction reward would be \(\frac{1}{5}\) and \(8\times 0.1\). Note that here we use a weight of 1 for all the rewards for clarity, which is different from the weights we used in the final model.}
  \label{main_flow}
\end{figure*}

\section{Methodology}

In this section, we introduce the methodology of Refine-POI, as shown in Figure \ref{main_flow}, which contains the following components: (i) topology-aware \acl{SID} and (ii) reinforcement fine-tuning. Topology-aware \acp{SID} serves as the base of our method and bridges the gap between raw data and LLMs. Recommendation-driven rewards are the core of our training, producing rewards based on LLMs’ outputs, which are then converted into policy gradients. 

\subsection{Topology-aware \acp{SID}}
\subsubsection{Feature extraction} We follow~\cite{10.1145/3711896.3736981} to extract features from the raw data. We map longitude and latitude into region IDs by mapping them into Google Maps' Plus Codes. We extract temporal features by allocating each POI visit into 24 time slots and use the top 10 time slots for a POI as its temporal feature. Furthermore, we assign the top 10 most frequent visitors for each POI as their user collaborative signals, which only considers the visits in training set. We construct one-hot vectors for each of the aforementioned features. The semantic representation of a POI is defined as:
\begin{equation}
p_e = concat(c,r,t,c_u),    
\end{equation}
where $p_e$ is the feature vector, $c$ represents the POI category, $r$ denotes the region, $t$ indicates the temporal feature, and $c_u$ contains the user collaborative signal.

\subsubsection{POI quantization}
Before quantization, we first pre-train an encoder with a contrastive objective to obtain the smoothed input embeddings. Given an input vector $p_{e}$, we generate augmented views via Gaussian noise injection, denoted as $\tilde{p}_{e} = p_{e} + \epsilon$. These are mapped to the latent space as $z_i = f_\theta(\tilde{p}_{e})$. The encoder is optimized using the InfoNCE loss \cite{oord2018representation} to maximize the similarity between positive pairs $(z_i, z_j)$ originating from the same $p_{e}$, while contrasting them against negatives:
\begin{equation}
    \mathcal{L}_{i,j} = - \log \frac{\exp(\text{sim}(z_i, z_j) / \tau)}{\sum_{k=1}^{2N} \mathbb{1}_{[k \neq i]} \exp(\text{sim}(z_i, z_k) / \tau)},
\end{equation}
where $\text{sim}(\cdot)$ denotes cosine similarity, $N$ is the batch size, and $\tau$ is the temperature parameter.
The encoder is frozen during the quantization stage. The input vector $p_e$ is then mapped into embeddings $\hat{p_e}$.

We propose a novel \acl{SID} quantization method. Our method focuses on preserving the semantic continuity. That is to say, POIs with close semantic ID indices should also be semantically similar. For example, a \acl{SID} with prefix 
$\langle A\_1, 2 \rangle$ $\langle B\_2, 3 \rangle$ $\langle C\_4, 5 \rangle$ $\langle D\_1, 1 \rangle$ 
is likely to share more semantic information with 
$\langle A\_1, 2 \rangle$ $\langle B\_2, 3 \rangle$ $\langle C\_4, 5 \rangle$ $\langle D\_1, 2 \rangle$ 
than with 
$\langle A\_1, 2 \rangle$ $\langle B\_2, 3 \rangle$ $\langle C\_4, 5 \rangle$ $\langle D\_8, 8 \rangle$.
To generate fine-grained \aclp{SID}, we employ a multi-stage quantization framework. We use a \emph{hierarchical self-organizing map} (HSOM) composed of $L$ layers. Unlike tree-structured approaches that partition the space into local sub-maps, our architecture employs a residual design where each layer consists of a single global SOM. This ensures that every prototype vector gets sufficient updates during the training.

Let $\hat{p_e} \in \mathbb{R}^d$ be the input embedding. The quantization proceeds sequentially. We define the input to the first layer as the original vector, $r^{(0)} = \hat{p_e}$. For each subsequent layer $l \in \{1, \dots, L\}$, the SOM quantizes the incoming vector and passes the quantization error (residual) to the next stage.

Formally, at layer $l$, the SOM identifies the discrete code $c^{(l)}$ by minimizing the distance between the input residual $r^{(l-1)}$ and the layer's codebook $W^{(l)} = \{w^{(l)}_1, \dots, w^{(l)}_K\}$:
\begin{equation}
    c^{(l)} = \operatorname*{arg\,min}_{k} \| r^{(l-1)} - w^{(l)}_k \|_2.
\end{equation}
After quantization, we compute the residual vector $r^{(l)}$:
\begin{equation}
    r^{(l)} = r^{(l-1)} - w^{(l)}_{c^{(l)}}.
\end{equation}
The final \acl{SID} is the concatenation of these hierarchical codes:
\begin{equation}
    \text{SID}(\hat{p_e}) = \langle 1, c^{(1)} \rangle \oplus \langle 2, c^{(2)} \rangle \oplus \dots \oplus \langle L, c^{(L)} \rangle.
\end{equation}
The hierarchical structure is optimized sequentially in a layer-by-layer fashion. The training process begins with the first layer ($l=1$) using the original input vectors. Once layer $l$ converges, its parameters are frozen, and the residuals $r^{(l)}$ are computed to serve as the training data for the subsequent layer $l+1$. This greedy approach ensures that each layer focuses exclusively on refining the quantization error left by its predecessors.
For a specific layer, we use a batch-based \ac{SOM} algorithm. The grid consists of a 2D lattice where each node $k$ possesses a fixed coordinate $u_k$ and a learnable prototype $w_k$. The grid consists of a 2D lattice where each node $k$ possesses a fixed coordinate $u_k$ and a learnable prototype $w_k$.
The weights are initialized from a scaled Gaussian distribution to ensure a compact starting state centered around the residual mean.

We employ a stable batch update rule rather than standard online learning. Given a batch of residuals $\mathcal{B} = \{r_1, \dots, r_B\}$, we first compute the best matching units (BMUs) $k^*_i$ for all inputs. The influence of a winner on its neighbor $k$ is determined by a Gaussian neighborhood function:
\begin{equation}
    h(k, k^*_i) = \exp\left( -\frac{\| u_k - u_{k^*_i} \|^2}{2\sigma^2} \right),
\end{equation}
where $\sigma$ is the neighborhood radius, and $u_k$ indicates the grid coordinates.

The prototypes are updated by aggregating the weighted difference vectors across the batch. To ensure stability, we normalize the update by the total accumulated neighborhood influence:
\begin{align}
    \Delta w_k & = \frac{\sum_{i=1}^{B} h(k, k^*_i) \cdot (r_i - w_k)}{\sum_{i=1}^{B} h(k, k^*_i) + \epsilon}
    \\
    w_k & \leftarrow w_k + \eta \cdot \Delta w_k,
\end{align}
where $\eta$ is the learning rate and $\epsilon$ is a small constant for numerical stability. Crucially, this normalization preserves semantic continuity. By smoothing the updates based on the local density of the batch ($\sum h$), we ensure that the topological ordering is maintained robustly against variance in the input stream. This guarantees that semantically similar residuals map to adjacent coordinates without the topological disruptions often observed in stochastic updates.

In practice, we use letters in alphabetical order as prefixes to represent layers for \acp{SID} and use coordinates to indicate the codebook vectors, such as $\langle A\_1,2 \rangle$ $\langle B\_2,3 \rangle$ $\langle C\_4,5 \rangle$ $\langle D\_1,1 \rangle$.

\subsection{Reinforcement fine-tuning}
The \ac{RFT} framework consists of (i) trajectory prompting and
(ii) recommendation-driven rewards.

\subsubsection{Trajectory prompting}
Trajectory prompting is proposed to make check-in trajectory data compatible with LLMs. It is a combination of long-term memory and short-term memory.

\textbf{Long-term memory.}
We extract historical trajectories to augment the current trajectory. Specifically, given a user’s entire data \(D_u\), we split \(D_u\) into several subsections proportional to its length, thus roughly mimicking a monthly data update.
We update the long-term memory when the incoming current trajectory reaches the next subsection, at which point a new subsection of historical trajectories is added to the existing long-term memory.

\textbf{Short-term memory.}
The short-term memory indicates the latest trajectory of the user for whom we want to make a recommendation. For a trajectory \(T_i^u(k)\), we use only \(T_i^i(k-1)\) as short-term memory, and the last entry of \(T_i^u(k)\) will be the target.

\textbf{Trajectory prompt generation.}
We construct the prompt in QA format. We follow the system prompt from DeepSeek-R1 \cite{guo2025deepseek}. For user prompt, we use element pairs `<history></history>' and `<current></current>' to indicate the long-term and short-term memory, respectively. Within the element pair, we describe each check-in record as a tuple (Check-in-Time, \ac{POI}-ID, \ac{POI}-Category, the distance between the current POI and the previous POI).

\subsubsection{Recommendation-driven reward}
There are two challenges in applying GRPO \cite{shao2024deepseekmath, guo2025deepseek} based \ac{RFT} to \ac{POI} recommendation: First, general R1-like models focus on tasks such as mathematics and coding. Their evaluation of answers is usually binary (for instance, whether the answer to a math question is correct or if the code can be executed). This differs from recommendation, where the output is often a list of items. We cannot simply evaluate whether a recommendation list is correct, because we have only one ground-truth item rather than the entire list. Second, LLMs tend to include duplicate items in their recommendation lists, especially after \ac{SFT}. To address these two challenges, we propose a recommendation-driven reward, which consists of a list format reward, a \ac{RR} reward, a soft accuracy reward, and a distinct reward.

\textbf{List format reward.} Extending the conventional syntax format reward, our list format reward ensures that the recommendation list contains the desired number of items. For each completion \(o_i\), in addition to checking the syntax format \(f_{\mathrm{syn}}\), we verify that the completion contains exactly \(k\) recommended items. We apply a binary reward scheme: completions with correct syntax and exactly \(k\) items receive a reward of 1; otherwise, they receive 0, as formulated below:
\begin{equation}
\mathit{reward}_{F}(o_i) =
\begin{cases}
1, & \text{if } f_{\mathrm{syn}} = 1 \wedge \#{\mathrm{item}} = k, \\
0, & \text{otherwise}.
\end{cases}
\end{equation}

\textbf{Reciprocal rank reward.} Since we have only a single ground-truth item rather than the full recommendation list, we can only measure whether the ground-truth item appears in the list. However, we can evaluate list quality by the rank of the ground-truth item, where rank is its position in the recommendation list. Inspired by the \ac{MRR} metric, which evaluates the mean reciprocal rank of the ground-truth item, we propose the \ac{RR} reward. As shown in Eq.~\ref{reward_rr}, we follow the \ac{MRR} design, assigning a reward equal to the reciprocal of the ground-truth item's rank:
\begin{equation}
\mathit{reward}_{RR}(o_i) =
\begin{cases}
\displaystyle \frac{1}{\mathit{rank}_i}, & 
\text{if } GT_i \in o_i \wedge \mathit{reward}_{F}(o_i) = 1, \\
0, & \text{otherwise}.
\end{cases}
\label{reward_rr}
\end{equation}

\textbf{Soft accuracy reward.} At the beginning of training, the \ac{RR} reward may not work because the model is still learning the format. Therefore, we introduce a soft accuracy reward that is more tolerant of format errors. As described in Eq.~\ref{reward_soft_acc}, a reward of 1 is assigned if the ground-truth item is in the answer and the completion has the correct syntax; otherwise, the reward is 0:
\begin{equation}
\mathit{reward}_\mathit{soft}(o_i) =
\begin{cases}
1, & 
\text{if } GT_i \in o_i \wedge f_{\mathrm{syn}} = 1, \\
0, & \text{otherwise}.
\end{cases}
\label{reward_soft_acc}
\end{equation}

\textbf{Distinction reward.} To encourage the model to recommend diverse items, we design a distinction reward, which measures the number of distinct items in the recommendation list. It is formulated as follows:
\begin{equation}
\mathit{reward}_\mathit{dis}(o_i) =
\begin{cases}
\lvert\{\text{item}_i\}\rvert, & 
\text{if } \mathit{reward}_{F}(o_i) = 1, \\
0, & \text{otherwise}.
\end{cases}
\label{reward_dis}
\end{equation}

\textbf{Length reward.} We notice that the model, during training, may fall into a shortcut where it outputs only a few words of the reasoning process. To improve the training stability, we use a length reward that is the ratio of output length to a hyperparameter target length, which should be no more than one, as follows:
\begin{equation}
\mathit{reward}_\mathit{len}(o_i) = \min\left(1, \frac{len_{output}}{len_{target}}\right).
\label{reward_len}
\end{equation}
The overall reward for each completion \(o_i\) is the weighted sum of the five rewards:
\begin{equation}
\begin{split}
\mathit{reward} = {}& w_1\cdot \mathit{reward}_{F} + w_2\cdot\mathit{reward}_{RR} +{}\\
& w_3\cdot
\mathit{reward}_\mathit{soft} + w_4\cdot\mathit{reward}_\mathit{dis}+ w_5\cdot\mathit{reward}_\mathit{len},
\end{split}
\end{equation}
where \(w_i\) is the corresponding weight.

\begin{table*}[htbp]
  \centering
  \caption{Comparison of methods on the NYC, TKY, and CA datasets. Improvement indicates the relative percentage increase of the best Refine-POI variant over the strongest baseline.}
  \label{tab:main}
  \setlength{\tabcolsep}{2.5pt}
  \resizebox{\textwidth}{!}{%
  \begin{tabular}{lll cccc cccc cccc}
    \toprule
    Method & Base LLM & Training paradigm&\multicolumn{4}{c}{NYC} & \multicolumn{4}{c}{TKY} & \multicolumn{4}{c}{CA} \\
    \cmidrule(lr){4-7} \cmidrule(lr){8-11} \cmidrule(lr){12-15}
           &          & & Acc@1 & Acc@5 & Acc@10 & MRR & Acc@1 & Acc@5 & Acc@10 & MRR & Acc@1 & Acc@5 & Acc@10 & MRR \\
    \midrule
    FPMC      & --& -- & 0.1003 & 0.2126 & 0.2970 & 0.1701 & 0.0814 & 0.2045 & 0.2746 & 0.1344 & 0.0383 & 0.0702 & 0.1179 & 0.0911 \\
    LSTM      & --& -- & 0.1305 & 0.2719 & 0.3283 & 0.1857 & 0.1335 & 0.2728 & 0.3277 & 0.1834 & 0.0665 & 0.1306 & 0.1784 & 0.1201 \\
    PRME      & --& -- & 0.1159 & 0.2236 & 0.3105 & 0.1712 & 0.1052 & 0.2278 & 0.2944 & 0.1786 & 0.0521 & 0.1034 & 0.1425 & 0.1002 \\
    STGCN     & --& -- & 0.1799 & 0.3425 & 0.4279 & 0.2788 & 0.1716 & 0.3453 & 0.3927 & 0.2504 & 0.0961 & 0.2097 & 0.2613 & 0.1712 \\
    PLSPL     & --& -- & 0.1917 & 0.3678 & 0.4523 & 0.2806 & 0.1889 & 0.3523 & 0.4150 & 0.2542 & 0.1072 & 0.2278 & 0.2995 & 0.1847 \\
    STAN      & --& -- & 0.2231 & 0.4582 & 0.5734 & 0.3253 & 0.1963 & 0.3798 & 0.4464 & 0.2852 & 0.1104 & 0.2348 & 0.3018 & 0.1869 \\
    GETNext   & --& -- & 0.2435 & 0.5089 & 0.6143 & 0.3621 & 0.2254 & 0.4417 & 0.5287 & 0.3262 & 0.1357 & 0.2852 & 0.3590 & 0.2103 \\
    STHGCN    & --& -- & 0.2734 & \underline{0.5361} & 0.6244 & 0.3915 & 0.2950 & \underline{0.5207} & \textbf{0.5980} & \underline{0.3986} & 0.1730 & \underline{0.3529} & \underline{0.4191} & 0.2558 \\
    \midrule
    LLM-Mob   & GPT-4o &Zero shot& 0.2343 & 0.5066 & 0.5660 & 0.3502 & 0.1545 & 0.3837 & 0.4451 & 0.2519 & 0.0293 & 0.0779 & 0.0908 & 0.0495 \\
    LLMMove   & GPT-4o &Zero shot& 0.1525 & 0.4429 & 0.5080 & 0.2690 & 0.1293 & 0.3601 & 0.4154 & 0.2248 & 0.074 & 0.1339 & 0.1607 & 0.1004 \\
    \midrule
    LLM4POI   & Llama3.1-8B     &SFT   & 0.3372 & 0.3982 & 0.5010 & 0.3807 & 0.3035 & 0.3797 & 0.4474 & 0.3492 & 0.2065 & 0.2586 & 0.3109 & 0.2371 \\
    GNPR-SID  & Llama3.1-8B  &SFT & \underline{0.3618} & 0.4472 & 0.5521 & \underline{0.4133} & \underline{0.3062} & 0.3930 & 0.4668 & 0.3662 & \underline{0.2465} & 0.2846 & 0.3384 & \underline{0.2559} \\
    \midrule
    Refine-POI   & Qwen2.5-7B &SFT& 0.3583 & 0.4255 & 0.5178 & 0.3905 & 0.3420 & 0.3996 & 0.4778 & 0.3754 & 0.2293 & 0.2896 & 0.3369 & 0.2593 \\ 
    Refine-POI   & Llama3.1-8B &SFT& \textbf{0.3751} & 0.4381 & 0.5269 & 0.4064 & \textbf{0.3552} & 0.4100 & 0.4723 & 0.3803 & \textbf{0.2514} & 0.3099 & 0.3478 & 0.2750 \\ 
    Refine-POI    & Qwen2.5-7B  &RFT& 0.3575 & 0.5899 & 0.6599 & 0.4592 & 0.3088 & 0.5113 & 0.5819 & 0.3961 & 0.2123 & 0.3741 & 0.4275 & 0.2812 \\ 
    Refine-POI   & Llama3.1-8B &RFT& 0.3638 & \textbf{0.6011} & \textbf{0.6696} & \textbf{0.4651} & 0.3214 & \textbf{0.5216} & \underline{0.5881} & \textbf{0.4072} & 0.2174 & \textbf{0.3755} & \textbf{0.4308} & \textbf{0.2833} \\ 
    \midrule
    Improvement & -- & -- & 3.67\% & 12.12\% & 7.24\% & 12.53\% & 16.00\% & 0.17\% & -1.66\% & 2.16\% & 1.98\% & 6.40\% & 2.79\% & 10.71\% \\    \bottomrule
  \end{tabular}
  }
\end{table*}

\section{Experiment}
\subsection{Experimental setup}
\subsubsection{Dataset}
We conduct experiments on three public datasets: Foursquare-NYC, Foursquare-TKY \cite{yang2014modeling}, and Gowala-CA \cite{cho2011friendship}. The first two datasets, collected over 11 months, comprise data from New York City and Tokyo, sourced from Foursquare. The Gowala-CA dataset, from the Gowalla platform, covers a broader geographical area and time period, encompassing California and Nevada. We use data that has been preprocessed as per the methods detailed by \citet{yan2023spatio}. The data is preprocessed as follows: 
\begin{enumerate*}[label=(\roman*)]
\item Filter out Points of Interest (\acp{POI}) with fewer than 10 visit records in history; 
\item Exclude users with fewer than 10 visit records in history; 
\item Divide user check-in records into several trajectories with 24-hour intervals, excluding trajectories that contain only one check-in record. 
\end{enumerate*}
The check-in records are also sorted chronologically: the first 80\% are used for the training set, the next 10\% are defined as the validation set, and the last 10\% are defined as the test set. Note that the validation and test set have to contain all users and \acp{POI} that appear in the training set. 
The unseen users and \acp{POI} would be removed from the validation and test set.
\subsubsection{Baselines}
We compare Refine-POI with 12 baselines, categorized into conventional sequence models, deep learning models, and LLM-based approaches.

\noindent\textbf{Conventional sequence models \& Deep learning models.} 
\textbf{FPMC} \cite{rendle2010factorizing} combines Markov chains with matrix factorization for sequential prediction. 
\textbf{LSTM} \cite{hochreiter1997long} captures long-term dependencies in sequential data. 
\textbf{PRME} \cite{feng2015personalized} learns sequential transitions via pairwise ranking metric embeddings. 
\textbf{STGCN} \cite{zhao2020spatio} uses gating mechanisms to model spatio-temporal intervals. 
\textbf{PLSPL} \cite{wu2020personalized} employs attention for short-term and parallel LSTMs for long-term preferences. 
\textbf{STAN} \cite{luo2021stan} aggregates spatio-temporal correlations via bi-layer attention. 
\textbf{GETNext} \cite{yang2022getnext} uses a global trajectory flow map and GCNs for next-\ac{POI} prediction. 
\textbf{STHGCN} \cite{yan2023spatio} constructs a hypergraph transformer to capture inter- and intra-user relations.

\noindent\textbf{LLM-based Models.} 
\textbf{LLM-Mob} \cite{wang2023would} uses prompting to capture long- and short-term mobility dependencies. 
\textbf{LLMMove} \cite{10605522} integrates user preferences and geo-distance into the prompt. 
\textbf{LLM4POI} \cite{li2024large} uses \ac{SFT} to formulate recommendation as a QA task. 
\textbf{GNPR-SID} \cite{10.1145/3711896.3736981} incorporates \aclp{SID} into an \ac{SFT}-based recommender.

\subsubsection{Evaluation metrics}
We use ACC@\(k\) and \ac{MRR} as our evaluation metrics, which are widely used for the next \ac{POI} recommendation task.
ACC@\(k\) looks at what proportion of the test items would have been retrieved with the top-\(k\) recommended list and can be formalized as:
\begin{equation}
    \text{Acc}@k = \frac{1}{m} \sum_{i=1}^{m} \mathbbm{1}(\text{rank} \leq k),
\end{equation}
where $\mathbbm{1}$ is the indicator function. Rank is the rank of the order of the correct prediction in the recommendation list. A larger value represents better performance.
MRR@\(k\) not only looks at the presence of the test item but also its rank in the recommended list. MRR@k can be defined as:
\begin{equation}
    MRR@k = \frac{1}{m} \sum_i^m\frac{1}{rank}.
\end{equation}

\subsubsection{Implementation details}
For the prompt-based baselines, we use GPT-4o \cite{hurst2024gpt} as the backbone model. For the fine-tuning approaches, we employ Qwen2.5-7B \cite{qwen2.5, qwen2.5-1m} and Llama3.1-8B \cite{dubey2024llama} as the base models. We train all fine-tuned variants for 3 epochs using a learning rate of $5e^{-5}$ with cosine scheduling.
Regarding the reinforcement learning alignment, we did limited hyperparameter tuning with reward weights in NYC and applied the same reward weights to TKY and CA, as the goal of the work is not ultimate performance. We classify the reward as accuracy-related reward, format reward, and length reward, where \ac{RR} reward, soft accuracy reward, and distinction reward are included in the accuracy-related reward. We assign $\alpha, \beta, \sigma$ to accuracy-related reward, format reward, and length reward, respectively, where $\alpha + \beta + \sigma = 1$. As a result, we assign weights 0.6, 0.2, and 0.4 to $\alpha, \beta, \sigma$, balancing the primary ranking objectives with structural constraints. And inside accuracy-related reward, we assign 0.7, 0.2, and 0.1 to \ac{RR} reward, soft accuracy reward, and distinction reward. We set the KL penalty coefficient to \(0.01\), the number of rollouts to 8, and temperature for rollouts to 1.
For the SOM-based quantization, we set the number of layers to 4, with hierarchical grid sizes configured as $4 \times 6$, $4 \times 6$, $8 \times 8$, and $8 \times 8$, respectively, for all three datasets. All experiments were conducted on a cluster equipped with Nvidia A100 and H200 GPUs.

\begin{figure*}[t]
    \centering
    \begin{minipage}[t]{0.485\linewidth}
    \begin{tcolorbox}[colback=blue!4, colframe=blue!55!black, title=\textbf{Grounded reasoning}]
        \small
        \textbf{[Task]:} given the user's history and current trajectory, predict the next POI at 2013-01-14 05:08.\\
        \rule{\linewidth}{0.4pt}
        \textbf{Reasoning ($\mathtt{<think>}$, condensed):}
        \begin{itemize}\setlength\itemsep{1pt}
            \item \textbf{Cites facts:} \textit{``the user visited the park 11 times and the coffee shop 7 times.''}
            \item \textbf{Temporal logic:} \textit{``visited the park at the same hour (06:20, 06:31, 07:20)\ldots a regular morning routine.''}
            \item \textbf{Constraint:} the target time (05:08) matches the morning park routine.
        \end{itemize}
        \rule{\linewidth}{0.4pt}
        \textbf{Prediction ($\mathtt{<answer>}$):} \texttt{[<A\_1\_0>, <A\_2\_5>, \ldots]}, the model emits a ranked list of semantic IDs, placing the Park's ID (\texttt{<A\_1\_0>}) first via the time-of-day match (categories given in text for readability).
    \end{tcolorbox}
    \end{minipage}\hfill
    \begin{minipage}[t]{0.485\linewidth}
    \begin{tcolorbox}[colback=red!4, colframe=red!55!black, title=\textbf{Vacuous reasoning}]
        \small
        \textbf{[Task]:} given the user's history and current trajectory, predict the next POI at 2013-01-16 20:05.\\
        \rule{\linewidth}{0.4pt}
        \textbf{Reasoning ($\mathtt{<think>}$, verbatim excerpt):}\\
        \textit{``To determine the POI the user will visit next, we need to analyze the provided data and look for any patterns. A possible approach would be to examine the history of previous visits, focusing on the frequency and recency of visits and the type of POI. Given that the input is large, we'll focus on identifying the most frequently visited POI and its category.''}
        \rule{\linewidth}{0.4pt}
        \textbf{Prediction ($\mathtt{<answer>}$):} \texttt{[<A\_3\_2>, <A\_0\_5>, \ldots]}, the model emits a ranked list of semantic IDs, yet its reasoning never names a single visit, time, count, or POI of this user; it only restates a generic procedure.
    \end{tcolorbox}
    \end{minipage}
    \caption{Two reasoning patterns from Refine-POI on NYC. For readability, the grounded trace (left) is condensed into its key steps, with the italicized fragments being verbatim quotes from the model; the bullet structure and labels are ours, not the model's original layout. The vacuous trace (right) is shown as a verbatim excerpt. The grounded trace cites the user's specific visits, counts, and times to reach its prediction; the vacuous trace recites a generic procedure without engaging any fact from the input. Full prompts and outputs are omitted for brevity.}
    \label{fig:reasoning_case}
\end{figure*}
\begin{figure}[t]
    \centering
    \includegraphics[width=0.99\linewidth]{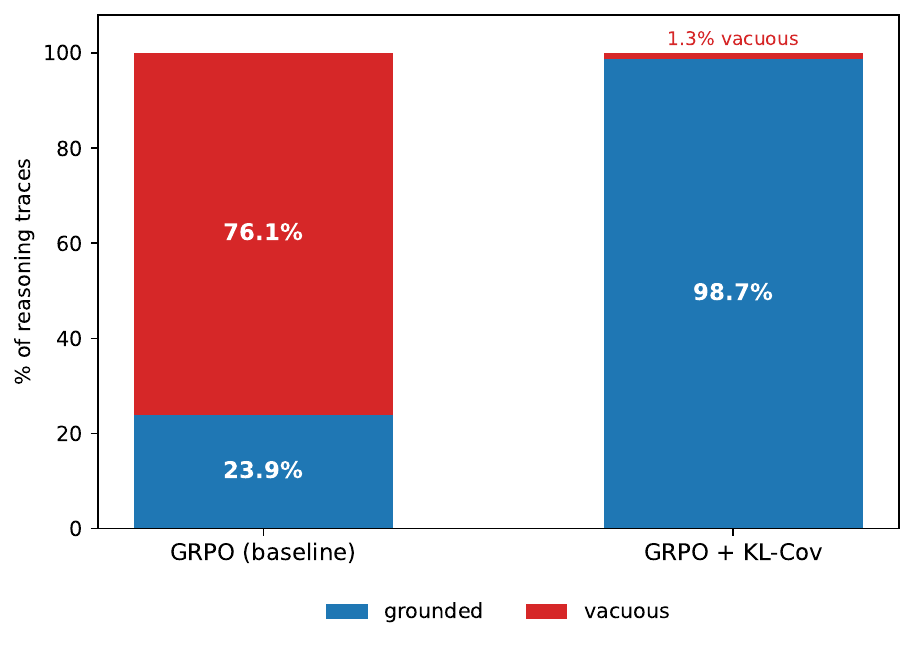}
    \caption{Reasoning analysis for Refine-POI on NYC: distribution of grounded vs.\ vacuous reasoning under standard GRPO and with KL-Cov entropy control. Entropy control raises the grounded fraction from $23.9\%$ to $98.7\%$.}
    \label{fig:reasoning_impact}
\end{figure}
\begin{figure}[t]
    \centering
    \includegraphics[width=0.99\linewidth]{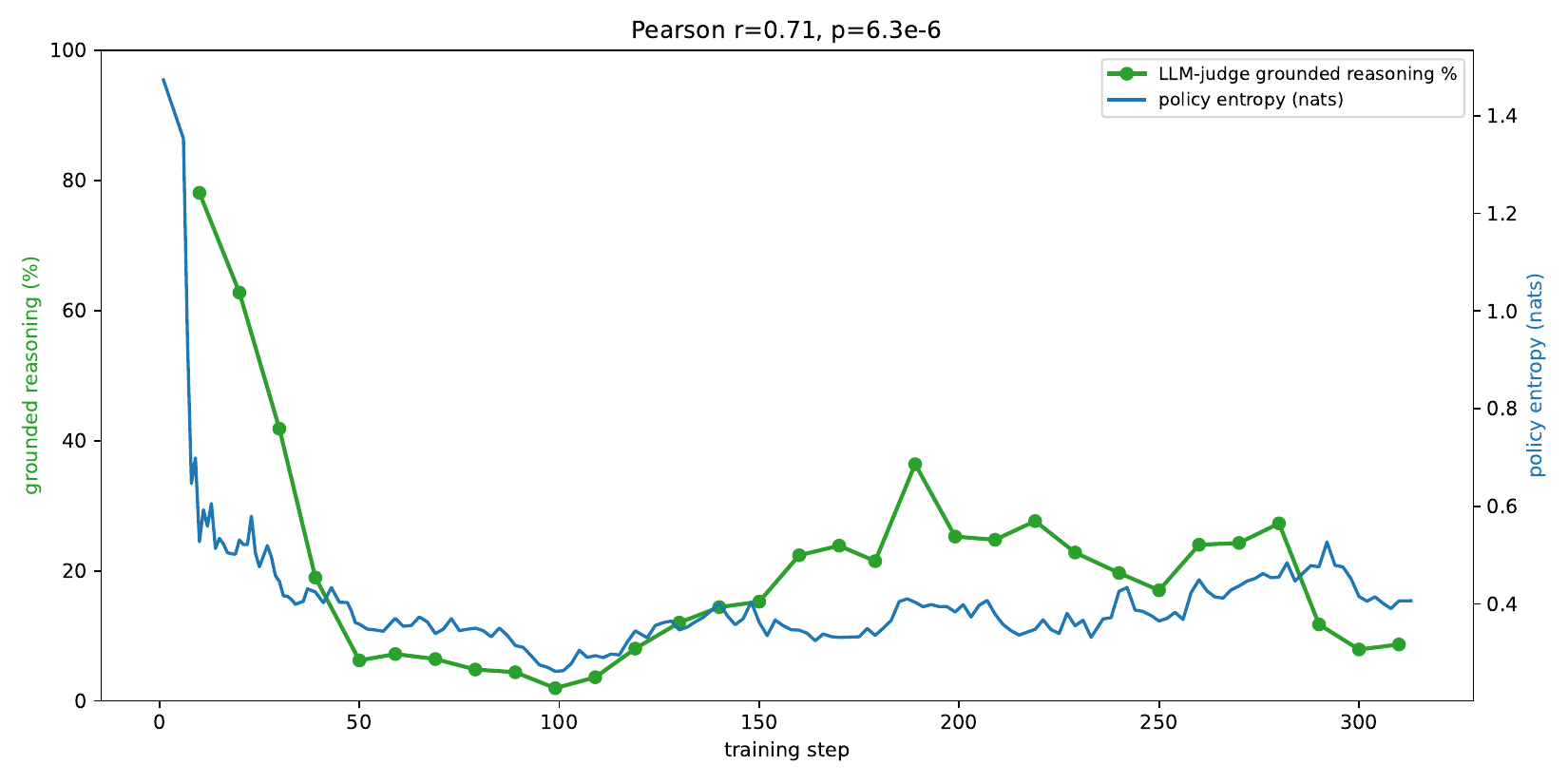}
    \caption{During GRPO training of Refine-POI on NYC, the fraction of grounded reasoning traces (green, LLM-judged) collapses in lockstep with the policy entropy (blue), reaching its minimum at the entropy trough (Pearson $r=0.72$).}
    \label{fig:entropy_grounding}
\end{figure}

\subsection{Main results}
Table \ref{tab:main} compares the performance of Refine-POI against baseline methods across all three datasets. Note that unless otherwise specified, ``Refine-POI'' refers to our primary configuration using Llama3.1-8B and the full \ac{RFT} framework.

\textbf{Performance analysis.} We observe a distinct performance trade-off between the training objectives. While the Refine-POI (\ac{SFT}) variant achieves the highest Acc@1, the Refine-POI (\ac{RFT}) variant consistently outperforms all methods in list-based metrics (Acc@5, Acc@10, and MRR). This dichotomy stems from the fundamental limitations of the objectives: \ac{SFT} optimizes for single-item prediction, maximizing the probability of the ground truth only. In contrast, Refine-POI uses \ac{RFT} to optimize the entire top-$k$ recommendation list, balancing accuracy with diversity. Consequently, Refine-POI provides robust, native list generation suitable for real-world scenarios, whereas \ac{SFT} models effectively ``overfit'' to the top-1 position at the expense of list quality.

\textbf{Limitations of \ac{SFT} inference.} It is important to address the methodological constraints of the baselines. Since \ac{SFT}-based models do not natively support ranked list generation, we employed a sampling-based inference strategy for comparison: generating 10 samples with high temperature ($T=1.0$), to maximize item distinction, and ranking them via the length-normalized average log-probability. Despite this computationally intensive workaround, the \ac{SFT} baselines fail to scale effectively on top-$k$ metrics (Acc@5, Acc@10). This empirically demonstrates that \ac{SFT} is insufficient for recommendation tasks requiring comprehensive candidate ranking, validating the necessity of our reward-guided optimization.

\subsection {Analysis and discussion}
\subsubsection{Reasoning}
A significant feature of R1-like models is their reasoning ability, which is attributed to the power of \ac{RFT} and outcome-based rule-based rewards~\citep{guo2025deepseek}. Since Refine-POI is also an \ac{RFT} framework, we naturally expect the model to exhibit reasoning capabilities for the next \ac{POI} recommendation task.
We observe that the generated reasoning processes diverge into two distinct patterns: \emph{grounded} or \emph{vacuous}. We define grounded reasoning as a process that explicitly cites facts from the provided history or contains inferences based on general knowledge. In contrast, we define vacuous reasoning as a process that relies on generic reasoning patterns without citing specific facts.
Figure~\ref{fig:reasoning_case} contrasts a grounded and a vacuous trace. In the grounded case, the model successfully finds a global pattern, that the user is a frequent visitor of the park and the coffee shop. It also provides quantitative evidence, noting that the user visited the park 11 times and the coffee shop 7 times in their history. The model links park visits to temporal information to infer a regular routine. As a result, the model identifies the association between the target time and the morning park routine, leading to an accurate prediction.
For vacuous reasoning, the model typically outputs generic analytical patterns, such as: ``To determine the POI that user 983 will visit next at 2013-01-16 20:05:07, we need to analyze the provided data and look for any patterns. A possible approach would be to examine the provided history of the user's previous visits to POIs, focusing on the frequency and recency of visits to the same POI. We may also consider the type of POI visited to determine any preferences. Given that the input is large, we'll focus on identifying the most frequently visited POI and its category.'' 
This indicates that the verbalized reasoning is post-hoc rather than functional: the model commits to a prediction via latent computation and emits a generic template only to satisfy the output format, rather than deriving the answer from the stated reasoning.

We analyze the reasoning traces of Refine-POI on NYC with an LLM-as-a-judge (using \texttt{gemini-3.1-flash-lite}) and find that, under standard GRPO the large majority of traces, $76.1\%$ on NYC, are vacuous rather than grounded. This high prevalence of vacuous reasoning prompts the question of what causes it; examining the training dynamics, we identify \emph{entropy collapse} during \ac{RFT} as the driver. The policy entropy collapses rapidly, falling from $1.47$ to $0.26$ nats within the first ${\sim}100$ update steps, and the reasoning quality collapses with it: the fraction of grounded traces starts high ($78\%$) and falls to near zero ($2\%$) precisely as the entropy reaches its minimum, with a Pearson correlation of $r=0.72$ between the grounded fraction and the policy entropy (Figure~\ref{fig:entropy_grounding}). Importantly, the entropy collapse \emph{precedes} the reasoning collapse, with entropy crashing within the first ten steps while the reasoning is still grounded, indicating that the loss of exploration is the cause rather than a symptom. The degradation is also persistent: it cannot be reversed at inference time. To test whether the collapse merely suppresses grounded reasoning at decoding or genuinely erodes the capability, we take the frozen collapsed policy and sweep its decoding temperature over $T \in \{0.5, 0.7, 1.0, 1.5, 2.0, 2.5\}$ (200 NYC prompts each). Re-injecting stochasticity does not help: the fraction of grounded traces never exceeds $4\%$ at any temperature (Table~\ref{tab:temp_probe}). At low temperature the model reproduces the same deterministic template, and at high temperature ($T \ge 2$) it degenerates into incoherent, repetitive text, with the output format collapsing entirely (the rate of well-formed answers drops to $0\%$) and self-similarity across prompts rising above $0.9$, rather than recovering grounded reasoning. This shows that entropy collapse erodes the policy's grounded-reasoning capability at the weight level, not merely its expression during decoding, which is why the collapse must be prevented \emph{during} training rather than patched afterwards.

\begin{table}[t]
\centering
\small
\caption{Decoding-temperature sweep on the frozen collapsed (GRPO) policy (200 NYC prompts per temperature). Re-injecting stochasticity at inference never restores grounded reasoning ($\le\!4\%$); at high temperature the output format breaks down and generations become near-duplicate (high self-similarity) rather than grounded. $T{=}1.0$ is the training temperature.}
\label{tab:temp_probe}
\begin{tabular}{lcccccc}
\toprule
Temperature $T$ & 0.5 & 0.7 & 1.0 & 1.5 & 2.0 & 2.5 \\
\midrule
Grounded (\%)      & 3.5 & 4.0 & 2.5 & 0.0 & 0.0 & 0.0 \\
Format-valid (\%)  & 100 & 100 & 99.5 & 99.5 & 0.0 & 0.0 \\
Self-similarity    & 0.37 & 0.29 & 0.18 & 0.54 & 0.92 & 0.96 \\
\bottomrule
\end{tabular}
\end{table}

To counteract entropy collapse, we regularize the policy entropy during \ac{RFT} using KL-Cov~\citep{cui2025entropy}. KL-Cov applies a KL penalty to the small fraction of tokens with the highest covariance between their log-probability and advantage---the over-confident updates that drive the collapse---thereby sustaining exploration without destabilizing training.

This intervention is effective: with KL-Cov, the policy maintains a high entropy and its reasoning remains grounded. On NYC, the fraction of grounded traces rises from $23.9\%$ under standard GRPO to $98.7\%$, while vacuous reasoning drops from $76.1\%$ to $1.3\%$ (Figure~\ref{fig:reasoning_impact}); task accuracy, meanwhile, remains essentially unchanged (Acc@1 $0.358 \!\to\! 0.375$, Acc@10 $0.669 \!\to\! 0.646$ on NYC). Entropy control thus transforms the model from one that recites generic templates into one that genuinely grounds its predictions in each user's specific history, yielding more faithful and interpretable recommendations.

This finding carries two broader implications. First, it identifies entropy collapse as a concrete, measurable failure mode of outcome-based \ac{RFT}: the model silently degenerates into reciting templates that ignore the input, yet this is invisible to standard accuracy metrics, which remain comparable. Monitoring the policy entropy (or the grounded-reasoning rate) is therefore a cheap and effective diagnostic for the health of the reasoning process. Second, it shows that the degradation is not inevitable: a lightweight entropy regularizer recovers faithful, fact-grounded reasoning at no accuracy cost, turning Refine-POI into a recommender whose justifications can be audited against the user's actual history. This makes the \emph{interpretability} of \ac{RFT}-based recommenders a controllable design choice rather than an emergent accident.

\subsubsection{User cold-start analysis}
The cold-start problem poses a challenge for next POI recommendation. Learning mobility patterns from users with little historical data is more difficult than from very active users. LLMs have strong generalization ability, and our location-aware trajectory prompt generation design incorporates users’ trajectories with POI locations, which may help LLMs learn generalized mobility patterns. To evaluate our method’s effectiveness with inactive users, we categorize users into inactive, normal, and very active groups based on the number of trajectories in the training set, designating the top 30\% as very active, the bottom 30\% as inactive, and the rest as normal.

We compare Refine-POI with STHGCN and LLM4POI, both of which are designed to address the cold-start problem and perform well on inactive users. Although our designs are not specifically for the cold-start problem, some of them, such as the user collaborative signal used to generate \aclp{SID}, provide additional information for inactive users. As shown in Table \ref{active user}, Refine-POI achieves the best performance for inactive users in the NYC dataset, while it is less competitive on the TKY dataset. Compared to LLM4POI, an \ac{SFT}-based model, Refine-POI has lower Acc@\(1\) on the TKY dataset. 

Overall, this comparison demonstrates the effectiveness of Refine-POI for inactive users. The major factor behind the performance difference between the NYC and TKY datasets may be the amount of data: TKY has much more data than NYC, so shared patterns among users become more critical. This suggests a potential design improvement for Refine-POI.

\begin{table}[ht]
\caption{User cold-start analysis on the NYC and TKY datasets.}
\label{active user}
\centering
\begin{tabular}{llccccc}
\toprule
User groups & Model & \multicolumn{2}{c}{NYC} & \multicolumn{2}{c}{TKY} \\ \cline{3-6} 
                     &                & Acc@1  & MRR   & Acc@1& MRR \\ \midrule
Inactive             & STHGCN        & 0.1460  &0.2247            &0.2164&0.3053\\
Normal               & STHGCN        & 0.3050  &0.4265            & 0.2659& 0.3596 \\
Very active          & STHGCN        & 0.3085   &0.4402      & 0.3464 &  0.4618\\ 
\midrule
Inactive             & LLM4POI           & 0.2500 &0.3043 & 0.2930  &   0.2703     \\
Normal               & LLM4POI           & 0.3689 &0.3839  &  0.3014 & 0.3195 \\
Very active          & LLM4POI           & 0.3711  &0.4079         &0.3011 & 0.4092    \\   
\midrule
Inactive             & Refine-POI           & 0.2832 & 0.3630 & 0.2655  &   0.3389     \\
Normal               & Refine-POI           & 0.3711 &0.4602  & 0.2814 & 0.3630 \\
Very active          & Refine-POI           & 0.3860  &0.6455        & 0.3768 & 0.4707        \\ \bottomrule
\end{tabular}
\end{table}

\subsubsection{\acp{SID} Semantic continuity analysis}

To illustrate the semantic continuity of our generated \aclp{SID} across different dimensionalities (1D vs.\ 2D), we adopt a null reference framework inspired by the gap statistic \cite{tibshirani2001estimating}. This approach evaluates the quality of semantic clusters by comparing the model's geometric properties against the expected properties of a uniform random distribution (null hypothesis). We define two complementary metrics under this framework:
\begin{enumerate}
    \item \textbf{\Ac{NICC}}: Following the original gap statistic formulation, we measure the tightness of semantic clusters. Similar to the gap statistic, we evaluate the validity of semantic clusters by comparing their intra-class dispersion $\sigma_{c}$ against the expected dispersion of a null reference distribution $\sigma_{random}$ (uniform noise) within the ID space:
    \begin{equation}
        \mathcal{R} = \frac{\sigma_{c}(\mathbf{z})}{\sigma_{random}(\mathbf{z})},
    \end{equation}
    where $\sigma_{c} = \frac{1}{|C|} \sum_{i \in C} d(\mathbf{z}_i, \mathbf{\mu}_c)$ is the mean centroid distance, and $\mathbf{z}$ represents the mapped semantic IDs. For 1D methods, this space is the linear integer domain $\mathbb{Z} \in [0, N]$; for 2D SOM methods, it is the topological grid coordinates $\mathbb{Z}^2 \in [H \times W]$. A lower value of $\mathcal{R}$ indicates that the model has successfully mapped semantically related POIs to contiguous regions in the ID codebook.
    \item \textbf{\Ac{NICS}}: Extending the comparison logic to global structure, we evaluate the distinctiveness of semantic categories by comparing their inter-class separation $\delta_{inter}$ against the expected separation of a null reference distribution $\delta_{random}$ (random centroids) within the same ID space:
    \begin{equation}
        \mathcal{S} = \frac{\delta_{inter}(\mathbf{\mu})}{\delta_{random}(\mathbf{\mu})},
    \end{equation}
    where $\delta_{inter} = \mathbb{E}_{c \neq c'}[d(\mathbf{\mu}_c, \mathbf{\mu}_{c'})]$ is the average pairwise Euclidean distance between the centroids of distinct categories. A higher value of $\mathcal{S}$ indicates that the model has successfully learned sharp boundaries between distinct semantic concepts, effectively using the codebook space to separate different functional regions.
\end{enumerate}

\begin{figure}[h]
    \centering
    \includegraphics[width=0.99\linewidth]{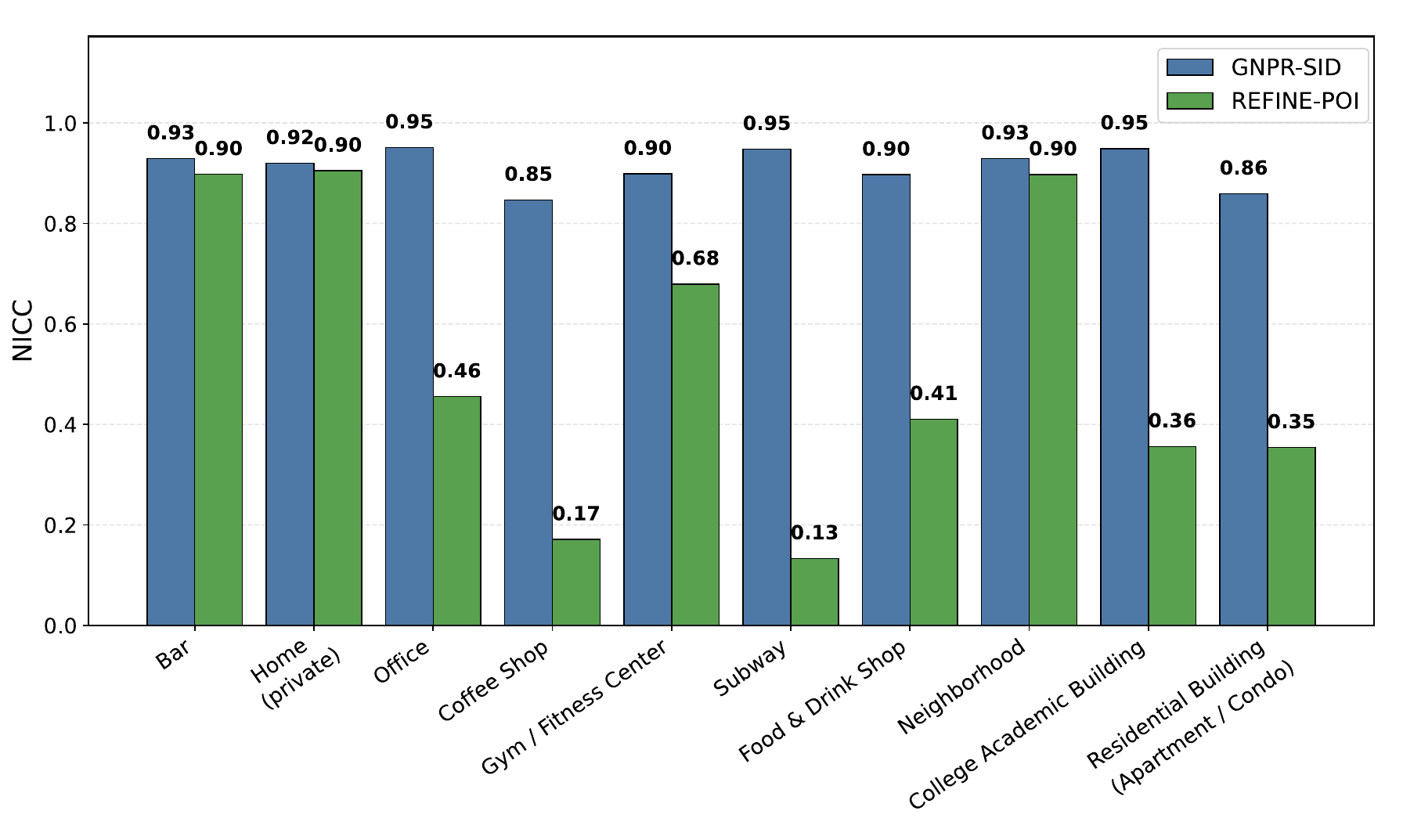}
    \caption{Comparison of normalized intra-class compactness on NYC between two types of \aclp{SID}.}
    \label{fig:cat_dist}
\end{figure}

\begin{figure}[h]
    \centering
    \includegraphics[width=0.99\linewidth]{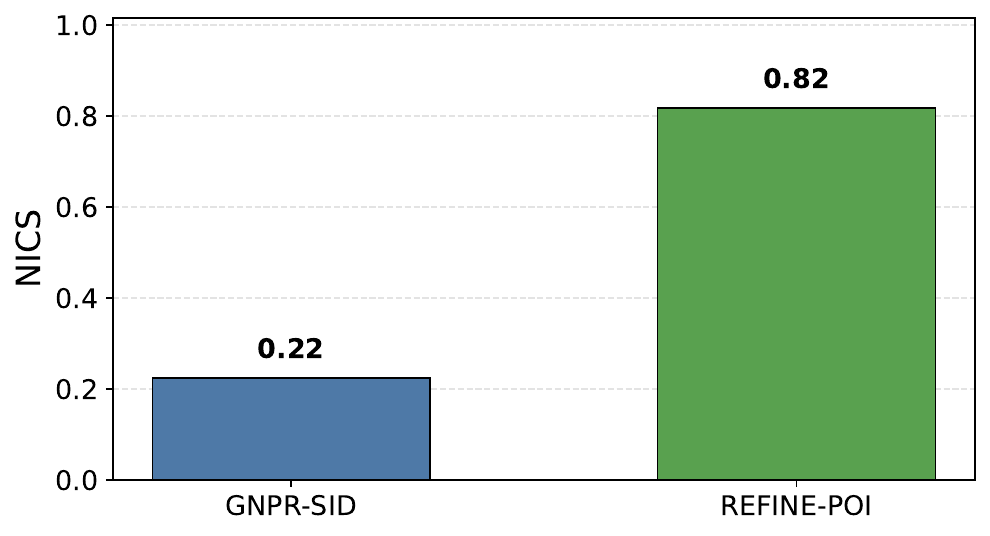}
    \caption{Comparison of normalized inter-class separation on NYC between two types of \aclp{SID}.}
    \label{fig:nics}
\end{figure}

\paragraph{Local and global analysis.}
In Figure \ref{fig:cat_dist} we compare the \ac{NICC} of the baseline GNPR-SID and our topology-aware \aclp{SID} on the NYC dataset for the top 10 categories. The results demonstrate that Refine-POI consistently achieves a lower dispersion ratio, reducing the intra-class spread by as much as 86\% relative to the baseline. This confirms that Refine-POI preserves semantic continuity, ensuring that semantically similar POIs are mapped to topologically neighboring codebook vectors.
Similarly, in Figure \ref{fig:nics}, we show a comparison of \ac{NICS} for the same categories. The results show that our inter-class separation is approximately 2.5 times higher than the baseline, indicating superior categorical distinctiveness.

\noindent%
Finally, Table \ref{tab:topology} presents the global average for both metrics. The results confirm a consistent structural advantage: Refine-POI achieves significantly tighter clusters (lower NICC) and sharper semantic boundaries (higher NICS) compared to the baseline across the entire dataset.

\begin{table}[t]
    \centering
    \caption{Global semantic continuity analysis. We report the \textbf{Global Average} \ac{NICC} and \ac{NICS}. \textbf{NICC ($\downarrow$)}: Lower is better (tighter clusters). \textbf{NICS ($\uparrow$)}: Higher is better (more distinct categories).}
    \label{tab:topology}
    \begin{tabular}{lcc}
        \toprule
        \textbf{Method} & \textbf{Avg. NICC} ($\downarrow$) & \textbf{Avg. NICS} ($\uparrow$) \\
        \midrule
        GNPR-SID (Baseline) & 0.7941 & 0.2817 \\
        \textbf{Refine-POI (Ours)} & \textbf{0.6076} & \textbf{0.7016} \\
        \bottomrule
    \end{tabular}
\end{table}

\begin{table}[htbp]
  \caption{Training time and memory usage of LLM4POI and Refine-POI using 8 A100 GPUs on the NYC dataset.}
  \label{tab:efficiency}
  \centering
  \begin{tabular}{l c c c}
    \toprule
    Model       & Time per epoch & \# training epochs  & Memory \\
    \midrule
    LLM4POI     & 3.1 h/epoch   & 3      & 512 GB \\
    Refine-POI  & 4.5 h/epoch   & 3   & 632 GB \\
    \bottomrule
  \end{tabular}
\end{table}

\begin{table}[t]
    \caption{Ablation study on three datasets.}
    \label{tab:ablation}
    \centering
    \setlength{\tabcolsep}{3.5pt}
    \begin{tabular}{lcccccc}
        \toprule
        Model & \multicolumn{2}{c}{NYC}& \multicolumn{2}{c}{TKY}& \multicolumn{2}{c}{CA} \\ 
        \cmidrule(r){2-3}
        \cmidrule(r){4-5}
        \cmidrule{6-7}
         & Acc@1&MRR& Acc@1&MRR& Acc@1&MRR \\
         \midrule
         Refine-POI&0.3638&0.4651&0.3214&0.4072&0.2174&0.2833\\
         w/o SID &0.3400&0.4454&0.3154&0.3986&0.1939&0.2647\\
         w/o rr &0.1749&0.3218&0.1373&0.2728&0.1032&0.1947\\
         w/o soft acc &0.3512&0.4502&0.3114&0.3648&0.2044&0.2722\\
         w/o distinct &0.3372&0.4416&0.2979&0.3857&0.2015&0.2678\\
         w/o len &0.3582&0.4617&0.3131&0.3944&0.1993&0.2699\\
         \bottomrule
    \end{tabular}
\end{table}
\subsubsection{Efficiency analysis}
Table \ref{tab:efficiency} compares the training efficiency of the full-parameter LLM4POI baseline against our proposed Refine-POI. We observe that Refine-POI incurs a higher computational overhead in both time and memory. This is structurally inherent to the \ac{RFT} framework, which necessitates: (i) multiple candidate outputs per input, to estimate the group advantage; and (ii) extended sequence lengths, as the model must generate intermediate reasoning chains (CoT) alongside the final prediction. While \ac{SFT} baselines are more computationally distinct, they lack the capacity for explicit reasoning. Therefore, we accept the increased training cost of Refine-POI as a necessary trade-off to achieve robust top-$k$ ranking capabilities and interpretability.

\subsection{Ablation study}
We disentangle the contributions of Refine-POI's core components: (i) the topology-aware \ac{SID} representation, and (ii) the recommenda\-tion-driven reward structure. To this end, we evaluate five ablation variants by selectively removing specific modules: w/o \ac{SID} (reverting to atomic IDs), and four reward-specific ablations (w/o rr, w/o soft acc, w/o distinct, w/o len). For the reward variants, we re-normalize the weights of the remaining terms to maintain gradient scale consistency. 

As detailed in Table \ref{tab:ablation}, the full Refine-POI model consistently achieves superior performance across all metrics, confirming the synergistic value of each component. Notably, the w/o rr variant exhibits the sharpest performance degradation, verifying that the reciprocal rank reward serves as the primary signal for rank-sensitive optimization, effectively forcing the model to prioritize the ground-truth item at the top of the list.

\section{Conclusion}
We proposed Refine-POI, the first \ac{RFT} framework for next \ac{POI} recommendation, pairing topology-aware \aclp{SID} for semantic continuity with recommendation-driven rewards that optimize top-\(k\) lists from a single ground-truth item. Across three real-world datasets it achieves state-of-the-art top-\(k\) performance, mitigates the cold-start problem, and improves semantic continuity, with an ablation confirming each component.

Beyond accuracy, we uncovered a reasoning failure mode: under naive GRPO the policy's entropy collapses early in training and its reasoning degrades from \emph{grounded} (citing the user's history) to \emph{vacuous} (generic templates that ignore the input). The collapse precedes the degradation and cannot be undone at inference time; a lightweight entropy regularizer (KL-Cov) prevents it, restoring grounded reasoning from \(24\%\) to \(99\%\) of traces at no accuracy cost. Monitoring policy entropy is thus a simple safeguard that makes explanation faithfulness a controllable property of \ac{RFT}-based recommenders.

More broadly, effective \ac{RFT} hinges on domain-appropriate reward design and on preserving exploration, i.e., policy entropy, which proves essential not only for training stability but for keeping reasoning grounded. Limitations remain: \ac{RFT} is computationally heavier than \ac{SFT}, and grounded and vacuous traces attain comparable accuracy here, so the verbalized reasoning is not yet a causal driver of the prediction. Aligning faithful reasoning with task performance, for example, through process-level supervision, is a promising direction for future work.


\clearpage
\bibliographystyle{ACM-Reference-Format}
\balance
\bibliography{references}

\end{document}